\documentclass[aps,pra,epsfigure,showpacs,twocolumn]{revtex4}
\usepackage{amsmath}
\usepackage{amstext}
\usepackage{latexsym}
\usepackage{epsfig}
\usepackage{graphicx}
\usepackage{amsfonts}
\usepackage{bm}
\usepackage{amssymb}

\newcommand{\ket}[1]{\left\vert#1\right\rangle}
\newcommand{\modul}[1]{\left\vert#1\right\vert}

\newcommand{\one}{\mbox{$1 \hspace{-1.0mm}  {\bf l}$}}

\begin{document}

\title{Entanglement generation in harmonic chains: tagging by squeezing}

\author{M. Paternostro$^{1}$, M. S. Kim$^{1}$, E. Park$^{1}$, and J. Lee$^{2}$}
\affiliation{$^{1}$School of Mathematics and Physics, The Queen's University,
Belfast BT7 1NN, United Kingdom\\
$^{2}$Department of Physics, Hanyang University, Seoul 133-791, South Korea}
\date{\today}

\begin{abstract}

We address the problem of spring-like coupling between bosons in an open chain configuration where the counter-rotating terms are explicitly included. We show that fruitful insight can be gained by decomposing the time-evolution operator of this problem into a pattern of linear-optics elements. This allows us to provide a clear picture of the effects of the counter-rotating terms in the important problem of long-haul entanglement distribution. The analytic control over the variance matrix of the state of the bosonic register allows us to track the dynamics of the entanglement. This helps in designing a global addressing scheme, complemented by a proper initialization of the register, which quantitatively improves the entanglement between the extremal oscillators in the chain, thus providing a strategy for feasible long distance entanglement distribution. 
\end{abstract}

\maketitle
\section{Introduction}

The development of reliable strategies for quantum communication and information transfer has gained, in these recent years, an increasing importance in the quantum information processing (QIP) panorama. The reliable implementation of a quantum channel for the exchange and distribution of information is indeed central in many potential QIP applications~\cite{varie0,varie1}. Intuitively, one could think about a scenario in which the quantum channel and the processing device are two different entities which have to be {\it interfaced} at the right time of a given protocol. This implies the ability of switching on and off the interfacing interaction with sufficient degree of accuracy and a reliable control at the single-qubit level, which are very demanding requirements in general. On the other hand, the idea of exploiting collective interactions of intrinsically multipartite systems, governed by external potentials which globally address the entire register, has very recently encountered the interest of the QIP community~\cite{sougato}. A global addressing scheme offers advantages in terms of controllability of the device and protection from the decoherence channels unavoidably opened by any sort of local external intervention. 

Inspired by the progresses performed in the design, coupling and management of bosonic nanostructures, which can behave quantum mechanically~\cite{buks}, important efforts have been produced in order to better understand the role that multipartite systems of coupled bosons have in the transfer and propagation of quantum information~\cite{plenio,iohelenmyung}. The application of global addressing techniques to systems of continuous variable (CV) bosonic systems is appealing for a less demanding implementation of CV quantum information processing. 

In this work, we re-consider the issue of entanglement generation in a chain of harmonic oscillators coupled through nearest-neighbor spring-like forces induced by an external potential which addressed the whole system. One of the points of interests in our analysis is the role played by counter-rotating terms (present in the interaction Hamiltonian) in the entanglement generation process. This point, anticipated by the studies in~\cite{plenio}, is analyzed here by a change of perspective. Instead of solving by brute force the dynamical equations ruling the evolution of the bosonic register, we look for a formal decomposition of the time evolution operator in terms of linear optics elements, following the successful route initiated in~\cite{iohelenmyung}. We believe that this alternative approach clarifies the entanglement dynamics within the register and provides a more transparent picture of the role of the counter-rotating terms in such a process. Entanglement is found to be always present if the counter-rotating terms are included in the interaction Hamiltonian. However, we find the degree of bipartite entanglement between the first and last oscillator to be very small (a feature which is evident, despite it has not been stressed, in the analyses in~\cite{plenio}). In order to quantitatively improve the  entanglement settled between the ends of an open chain, we design a strategy based on proper initialization of the register (performed by locally acting on the state of the extremal oscillators only) and global addressing, following the same lines depicted in quantum state transfer protocols~\cite{qst}. We show how, physically, this improvement is possible because of the symmetry properties of the bosonic system. 

The reminder of the manuscript is organized as follows. In Section~\ref{modello} we introduce the interaction model here at hand. We discuss the technical tools used in order to derive effective decompositions of the time-evolution operator into linear optics operations. Effective all-optical setups can be introduced, which provide a visual picture of the evolution of an $N$-element register and we give an explicit example for a simple case. In Section~\ref{entanglement}, the entanglement generated in an open chain is quantified by means of the corresponding equivalent decompositions. We show that, as long as only the quantum correlations generated by the counter-rotating terms alone are considered, {\it end-to-end} entanglement in the chain is not favoured. Strong quantum correlations, which never disappear, are found between the first and the second oscillator in the chain. On the other hand, the entanglement between the first and the last oscillator is always very weak. A transparent physical interpretation of the time delay with which entanglement appears in the first-last oscillators subsystem is possible through the analysis of the corresponding all-optical setup. Section~\ref{improving} addresses a way to improve the results discussed in Section~\ref{entanglement}. By considering the physical system as a fictitious two-terminal quantum black box, we show that simple local pre-squeezing of the first and last element of the channel allows us to obtain several interesting effects. The end-to-end degree of entanglement can be quantitatively improved and any other bipartite $1\rightarrow{j}$ quantum correlation ($j=2,..,N-1$) can be correspondingly suppressed. 

\section{The model and the effective decomposition} 
\label{modello}
 
\begin{figure} [t]
\centerline{\psfig{figure=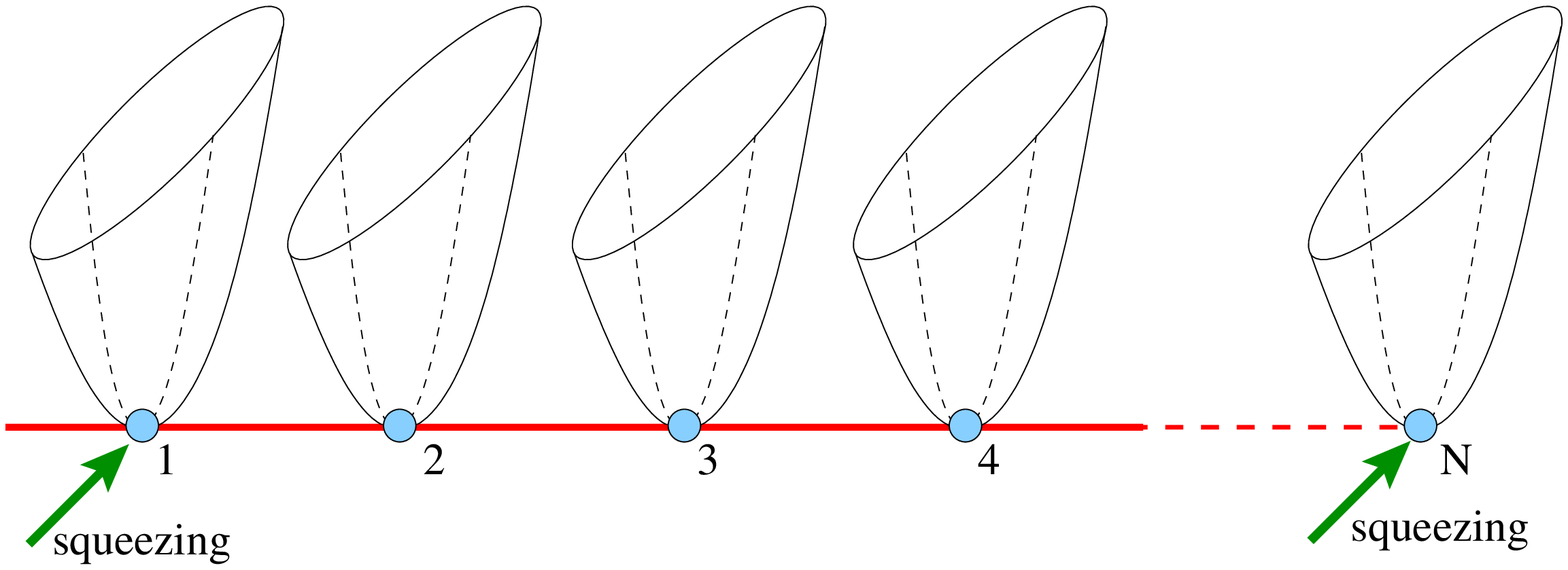,width=8.0cm,height=2.5cm}}
\caption{(color online) Diagrammatic scheme of the coupling configuration in Eq.~(\ref{interazione}). We consider $N$ harmonic oscillators, in an open linear chain, coupled through nearest-neighbor interactions. The squeezing operations performed onto oscillator $1$ and $N$ are part of the {\it tagging by squeezing} scheme suggested in Section~\ref{improving}.}  
\label{chain}
\end{figure}

We consider $N$ oscillators labelled by $j\in[1,N]$ and arranged in an open linear chain. The coupling between the oscillators is provided by a nearest-neighbor spring-like force settled by an external potential. By including the free dynamics of each harmonic oscillator, the corresponding Hamiltonian reads
\begin{equation}
\label{interazione}
\hat{H}_{chain}=\frac{\omega}{2}\sum^{N-1}_{j=1}\left(\hat{q}^{2}_{j}+\hat{p}^{2}_{j}\right)+\kappa\sum^{N-1}_{i=1}\hat{q}_{j}\hat{q}_{j+1},\hskip0.5cm(\hbar=1)  
\end{equation}
with $\hat{q}_{j}=(\hat{b}_{j}+\hat{b}^{\dagger}_{j})/\sqrt{2}$ and $\hat{p}_{j}=i(\hat{b}^{\dag}_{j}-\hat{b}_{j})/\sqrt{2}$ the position and momentum quadrature operators of the $j$-th oscillator respectively and $\hat{b}_{j}$ ($\hat{b}^{\dagger}_{j}$) the corresponding annihilation (creation) operator~\cite{barnett}. The coupling rates $\kappa$'s are real and time-independent. A sketch of the interaction configuration is provided in Fig.~\ref{chain}.
The form of the coupling terms deserves some comments as it is straightforward to see that each $\kappa\hat{q}_{j}\hat{q}_{j+1}$ in Eq.~(\ref{interazione}), expressed by means of the annihilation and creation operators, reads
\begin{equation}
\label{esplicito}
\kappa\hat{q}_{j}\hat{q}_{j+1}=\frac{\kappa}{2}(\hat{b}^{}_{j}\hat{b}^{}_{j+1}+\hat{b}^{}_{j}\hat{b}^{\dag}_{j+1}+\hat{b}^{\dag}_{j}\hat{b}^{}_{j+1}+\hat{b}^{\dag}_{j}\hat{b}^{\dag}_{j+1}).
\end{equation}
Eq.~(\ref{esplicito}) includes co-rotating terms ($\hat{b}_{j}\hat{b}^{\dag}_{j+1}+h.c.$) as well as counter-rotating (CR) terms ($\hat{b}_{j}\hat{b}_{j+1}+h.c.$)~\cite{barnett}. 

In this paper, we treat the CR terms on the same footage as the co-rotating ones, analyzing their relevance in entanglement generation processes in interacting bosonic systems. In order to analyze the time-evolution of the chain, we look for an effective decomposition of the propagator $\hat{U}(t)=e^{-i\hat{H}_{chain}t}$ in terms of linear-optics elements. 
We order the quadrature-operators as $\hat{\bf x}=(\hat{q}_{1},\hat{q}_{2},..,\hat{q}_{N},\hat{p}_{1},\hat{p}_{2},..,\hat{p}_{N})^{T}$ and divide $\hat{H}_{chain}$ as $\hat{H}_{chain}=\hat{H}^{p}_{chain}+\hat{H}^{q}_{chain}$, where $\hat{H}^{q}_{chain}$ ($\hat{H}^{p}_{chain}$) involves only the $q$-part ($p$-part) of $\hat{\bf x}$. In matrix form
\begin{equation}
\label{trid}
\hat{H}^{q}_{chain}=\frac{1}{2}
\begin{pmatrix}
\omega&\kappa&0&0&\cdots&0&0\\
\kappa&\omega&\kappa&0&\ddots&0&0\\
0&\kappa&\omega&\kappa&\ddots&0&0\\
\vdots&\ddots&\ddots&\ddots&\ddots&0&0\\
\vdots&\ddots&\ddots&\ddots&\ddots&\omega&\kappa\\
0&0&\cdots&\cdots&\cdots&\kappa&\omega\\
\end{pmatrix}.
\end{equation}
This is a tridiagonal matrix whose dimension depends on the number of elements in the register. The formal diagonalization of $\hat{H}^{q}_{chain}$ guides us in expressing the $q$-part of Eq.~(\ref{interazione}) in a picture defined by eigen-operators which are linear superpositions of $\hat{q}_{j}$'s. On the other hand, $\hat{H}^{p}_{chain}$ is already diagonal in the $\hat{\bf x}$ basis and its form is not changed by orthogonal transformations. Therefore, we discard this part of Eq.~(\ref{interazione}) from our explicit analysis and will include it only when necessary.

The simple form of Eq.~(\ref{trid}) allows for an efficient diagonalization~\cite{tridiagonal}, which helps us in identifying a proper pattern of coupling operations for the decomposition of $\hat{U}$. In order to keep our analysis general, we will refer to the well-known beam-splitter (BS) operator $BS_{jk}(\theta)=\exp[i\theta(\hat{q}_{j}\hat{p}_{k}-\hat{p}_{j}\hat{q}_{k})]$ of its reflectivity $\sin^2{\theta}$ as a {\it coupler} operator because this term can be used for both optical fields and mechanical oscillators. In the eigen-operator basis we write 
\begin{equation}
\label{formale}
\hat{H}^{q,N}_{chain}=\sum^{N}_{j=1}{E}^{N}_{j}(\hat{O}^{N}_{j})^2,
\end{equation}
where ${E}^{N}_{j}$'s are the eigen-frequencies of Eq.~(\ref{trid}) and $\hat{O}^{N}_{j}=\sum^{N}_{k=1}\alpha^{N}_{jk}\hat{q}_{k}$ ($j=1,..,N$) are the corresponding eigen-operators, expressed as normalized superpositions of the $\hat{q}_{k}$ quadratures with coefficient $\alpha^{N}_{jk}$. The set $\{{E}^{N}_{j},\hat{O}^{N}_{j}\}$ is parameterized by the dimension $N$ of the chain. As an explicit example, we consider the first non-trivial case represented by an open chain of $N=3$ where we have $\{{E}^{3}_{1},{E}^{3}_{2},{E}^{3}_{3}\}=\{{\omega}/{2},({\omega+\sqrt{2}\kappa})/{2},({\omega-\sqrt{2}\kappa})/{2}\}$. We introduce the matrix of coefficients $\alpha^{3}_{jk}$  
\begin{equation}
\label{coefficienti}
\alpha^{3}=\frac{1}{\sqrt{2}}
\begin{pmatrix}
1&0&-{1}\\
\frac{1}{\sqrt 2}&+{1}&\frac{1}{\sqrt 2}\\
\frac{1}{\sqrt 2}&-{1}&\frac{1}{\sqrt 2}\\
\end{pmatrix}.
\end{equation}
Thus, the spectrum of $\hat{H}^{q,3}_{chain}$ is a ladder, symmetric with respect to the {\it bare} eigen-frequency $\omega/2$. This is a general result for an odd number of oscillators: by increasing the dimension of the register, the spectrum of $\hat{H}^{q,2l+1}_{chain}$ (${l}\in{\mathbb Z}$) is never degenerate ans consists of $l$ different pairs of frequencies symmetrically shifted with respect to $\omega/2$.

Coming back to our example, the structure of the eigen-operators $\hat{O}^{3}_{j}$ ($j\in[1,3]$) turns out to be very informative in the research for a set of operations which can be used, starting from Eq.~(\ref{trid}), in order to get the diagonal form (\ref{formale}). Indeed, Eq.~(\ref{coefficienti}), suggests that a $50:50$ coupler operator $BS_{13}$
simplifies the structure of the coupling terms leaving us with oscillator $1$ being decoupled from the dynamics of the rest of the register. Then, a $50:50$ $\hat{B}_{23}$ operation will complete the diagonalization of $\hat{H}^{q}_{chain}$:   
\begin{equation}
\hat{H}^{q}_{chain}\stackrel{\hat{B}_{23}\hat{B}_{13}}{\longrightarrow}\frac{\omega}{2}\hat{q}^{''2}_{1}+\frac{\omega-\sqrt{2}\kappa}{2}\hat{q}^{''2}_{3}+\frac{\omega+\sqrt{2}\kappa}{2}\hat{q}^{''2}_{2}
\end{equation}
with $\hat{q}^{''}_{j}$ which are the new quadratures after BS's to be put in correspondence with $O^{3}_{j}$'s. The matching with Eq.~(\ref{formale}) is evident. Thus, the explicit diagonalization procedure of the $q$-part of the chain's Hamiltonian gives us information about the pattern of BS operations which have to be applied to the bare expression Eq.~(\ref{trid}) in order to get Eq.~(\ref{formale}). 
\begin{figure} [b]
\hspace*{-0.5cm}{\bf (a)}\hskip2.0cm{\bf (b)}\hskip2.9cm{\bf (c)}
\centerline{\psfig{figure=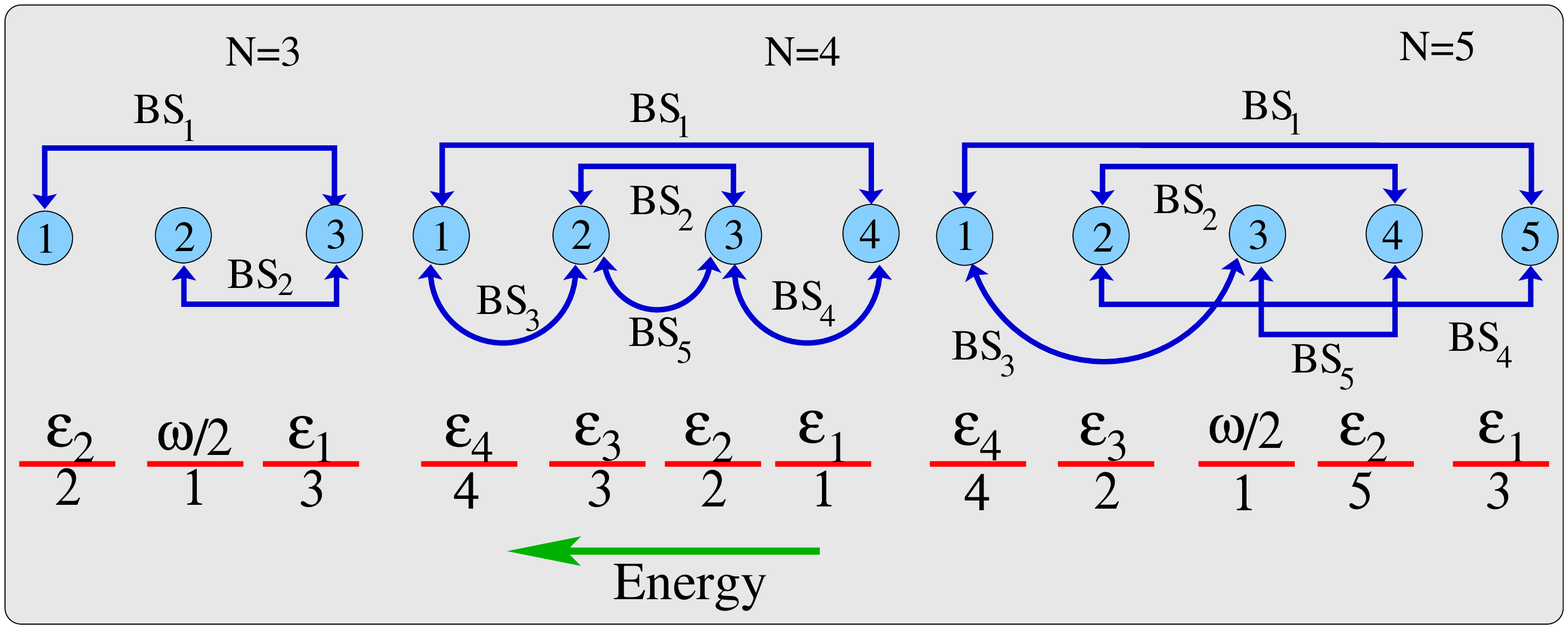,width=8.5cm,height=3.8cm}}
\caption{(color online) Pattern of coupler operators (or BS operators) for the diagonalization of $\hat{H}_{chain}$ for $N=3,4,5$ oscillators (panels {\bf (a), (b)} and ${\bf (c)}$, respectively). Straight (curved) lines denote $50:50$ (unbalanced) BS's. The time-order of the BS operations is such that BS$_{i}$ precedes BS$_{j}$ if $i<j$. The bottom part of the figure shows the correspondences between oscillators and eigen-frequencies induced by the corresponding BS's pattern.}
\label{pattern}
\end{figure}
In Fig.~\ref{pattern} we provide the sequence of BS's to apply in order to diagonalize the interaction Hamiltonian for $N=3,\,4$ and $5$ (panels ${\bf (a)}$, ${\bf (b)}$ and ${\bf (c)}$ respectively). Straight lines represent $50:50$ BS's, while curved ones stand for unbalanced BS's.  Some comments are in order. First, there is a striking difference between the even and odd number of oscillators. For the odd number case, up to $N=7$, a single unbalanced BS is required in order to diagonalize $H^{q}_{chain}$. On the other hand, the BS pattern for the even number case appears to be more complicated already for $N=4$, which is the first non-trivial configuration with even $N$, involving two unbalanced BS's between the pairs of oscillators $(1,2)$ and $(3,4)$. A second important difference between chains of opposite parity will be highlighted later on. We stress that there might be other, inequivalent ordering of coupler operations which diagonalize the interaction Hamiltonian. The choices presented in Figs.~\ref{pattern} ${\bf (a)}$ and ${\bf (c)}$ allow us to write the interaction part of the chain Hamiltonian in the special diagonal form corresponding to the association between oscillators and eigen-frequencies shown in the bottom part of Fig.~\ref{pattern}. There, the set $\{E^{N}_{j}\}$ is written as $\{\omega/2,{\cal E}^{N}_{1},..,{\cal E}^{N}_{N-1}\}$ with the subset $\{{\cal E}^{N}_{j}\}$ arranged in increasing order of frequencies (with $\kappa>0$)~\cite{commento}.
 
\begin{figure} [t]
\centerline{\psfig{figure=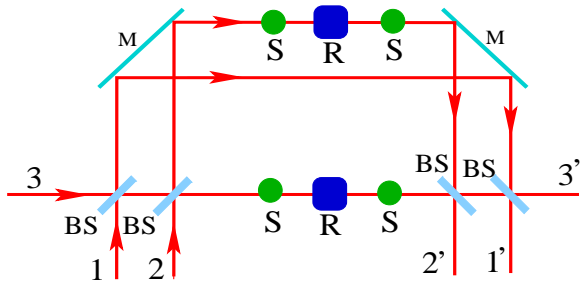,width=7.7cm,height=3.7cm}}
\caption{(color online) Equivalent interferometric setup corresponding to the decomposition of the time-evolution operator for an open chain of $N=3$ oscillators. $BS$ stands for beam-splitter ({\it i.e.} a coupler), $R$ indicates rotation and $S$ squeezing. $M$ stands for a mirror.}  
\label{3oscillatori}
\end{figure}

After the action of the collective coupler operator $\hat{B}_{coll}$, which collects the pattern (for a given $N$) discussed above, the total Hamiltonian of the chain reads $\hat{H}^{q}_{chain}=(\omega/2)[(\hat{O}^{N}_{1})^2+\sum^{N}_{j=1}(\hat{P}^{N}_{j})^2]+\sum^{N-1}_{j=1}{\cal E}^{N}_{j}(\hat{O}^{N}_{j})^2$. Here, $\{\hat{P}^{N}_{j}\}$ is the new set of momentum quadrature operators determined by the application of the coupler operations to $\{\hat{p}_{j}\}$. By specializing the discussion to the odd number of oscillators, the next step in order to find the decomposition of $\hat{U}(t)$ is the introduction of proper operations (acting on the elements of the register, oscillator $1$ exluded) which balance the differences between $\omega/2$ and ${\cal E}^{N}_{j}$. In the case of even $N$, these operations would involve the entire set of oscillators, without exclusions, as the eigen-spectrum of $\hat{H}^{q,2l}_{chain}$ ($l\in{\mathbb Z}$) does not include the bare frequency $\omega/2$. Conceptually, this balancing is an important step as it would allow us to look at the register as a set of new non-interacting harmonic oscillators. It is immediate to recognize that this is possible through the use of single-oscillator squeezing $\hat{S}_{j}(s_{j})=\exp[\frac{i}{2}s_{j}(\hat{O}^{N}_{j}\hat{P}^{N}_{j}+\hat{P}^{N}_{j}\hat{O}^{N}_{j})]$, which realizes $\hat{O}^{N}_{j}\rightarrow{e}^{-s_{j}}\hat{O}^{N}_{j},\,\hat{P}^{N}_{j}\rightarrow{e}^{s_{j}}\hat{P}^{N}_{j}$. We can thus formally write $\hat{S}^{\dag}_{coll}\hat{B}^{\dag}_{coll}\hat{U}(t)\hat{B}_{coll}\hat{S}_{coll}=\hat{R}_{coll}(t)\equiv\otimes^{N}_{j=2}\hat{R}_{j}(\phi_{j}(t))$, where $\hat{R}_{j}(\phi_{j})=e^{i\phi_{j}[(\hat{O}^{N}_{j})^{2}+(\hat{P}^{N}_{j})^{2}]}$ is the phase-space representation of a rotation operator of its angle $\phi_{j}$ and $\hat{S}_{coll}=\otimes^{N}_{j=2}\hat{S}_{j}(s_{j})$. By inverting the above relation, we find the formal expression 
\begin{equation}
\label{decomposizione}
\hat{U}(t)=\hat{B}_{coll}\hat{S}_{coll}\hat{R}_{coll}(t)\hat{S}^{\dag}_{coll}\hat{B}^{\dag}_{coll}.
\end{equation}
Analogously to the case of squeezing, the formal collective rotation involves all the oscillators but the one associated to the bare eigen-frequency $\omega/2$, which again is a specific feature of the odd $N$ case. Moreover, the rotations $\hat{R}_{j}(\phi_{j}(t))$'s contain the entire time dependence of $\hat{U}(t)$. The balancing induced by the squeezing operations imposes, in general, time-independent conditions as it relates the squeezing parameters $s_{j}$'s to the elements of the set $\{{\cal E}^{N}_{j}\}$. On the contrary, the formal identification of $\hat{S}^{\dag}_{coll}\hat{B}^{\dag}_{coll}\hat{U}(t)\hat{B}_{coll}\hat{S}_{coll}$ with $\hat{R}_{coll}$ imposes that the rotation angles $\phi_{j}$'s carry an explicit time dependence. As an example, we consider again $N=3$, where we have that 
\begin{equation}
\label{setparametri}
s_{2,3}=\frac{1}{4}\ln({{2{\cal E}^{3}_{2,1}}/{\omega}}),\hskip0.3cm\phi_{2,3}=\frac{t}{2}\sqrt{2{\cal E}^{3}_{2,1}\omega}.
\end{equation}
The decomposition Eq.~(\ref{decomposizione}) is a central result of our study. It allows us to provide a clear physical picture of the dynamics occurring within the linear chain, without explicitly solving the dynamical equations of motion of the oscillators~\cite{plenio}. Indeed, once the explicit form of $\hat{B}_{coll}$ is found, one can straighforwardly infer the evolution of the oscillators configuration simply by considering proper squeezing and rotations. This is equivalent to designing formal interferometric setups which could be used for proof-of-principle experiments where, at least for a few elements, the effects of CR terms could be simulated and observed. Motivated by these arguments, in Fig.~\ref{3oscillatori} we show the equivalent interferometer for $N=3$.
By inspection, we see that this equivalent configuration results in {\it concatenated} Mach-Zehnder interferometers where the oscillators involved are subject to different squeezing and rotation operations. In going from $N=3$ to $N=5$, the overall concatenated structure of the setup is preserved, with just more oscillators being involved. This is not the case for $N=4$, whose equivalent all-optical setup turns out to be more complicated than the one corresponding to $N=5$ for instance, with squeezing and rotations involving the entire register, as already stressed. The intrinsic difference between the even and odd $N$ cases should now be more evident. The most relevant discrepancy is caused by the absence of the bare frequency $\omega/2$ from the spectrum of $\hat{H}^{q,2l}_{chain}$. 

The second issue which has to be discussed here in relation to Eq.~(\ref{decomposizione}) is the role of the CR terms. It is easy to be convinced that the version of $\hat{H}_{chain}$ where the CR terms are excluded would have a contribution having the form $\kappa\sum^{N-1}_{j=1}\hat{p}_{j}\hat{p}_{j+1}$. Now the $p$-part of the chain Hamiltonian is also non-diagonal, with the same tridiagonal structure in the quadrature operators basis discussed in Eq.~(\ref{trid}). Therefore, the orthogonal transformation which diagonalizes $\hat{H}^{q}_{chain}$ (and the corresponding pattern of BS's) can be used in order to reduce the $p$-part as well, getting the same set of eigen-frequencies. The corresponding eigen-operators are superpositions of just the $\hat{p}_{j}$ quadrature operators with the same numerical coefficients $\alpha^{N}_{jk}$'s appearing in $\hat{O}^{N}_{j}$'s. This implies that, for the odd $N$ case, after the application of $\hat{B}_{coll}$, we end up with $\hat{B}^{\dag}_{coll}\hat{H}_{chain}\hat{B}_{coll}=(\omega/2)[(\hat{O}^{N}_{1})^2+(\hat{P}^{N}_{1})^2]+\sum^{N}_{j=2}{\cal E}_{j}[(\hat{O}^{N}_{j})^2+(\hat{P}^{N}_{j})^2]$, where $\hat{P}^{N}_{j}=\sum^{N}_{k=1}\alpha^{N}_{jk}\hat{p}_{k}$. Evidently, no squeezing is required in this case as the $q$ and $p$-parts of the Hamiltonian are already balanced by the diagonalization procedure. Thus, the corresponding time-evolution operation could be immediately reinterpreted as the tensorial product of formal rotation operators, one for each oscillator, showing that in this case the interferometric configurations sketched above are still valid: we only need to remove the squeezing operations. 

It should be clear, up to this stage, that the exchange of any information encoded in the elements of the bosonic register occurs entirely by means of the effective collective operation $\hat{B}_{coll}$. The remainder of the decomposition we have found, indeed, involves single-element operations which do not mutually mix the oscillators. Thus, by considering co-rotating terms only, we can see that the structure of $\hat{B}_{coll}$, for a given $N$, is unchanged.
This observation paves the way to the following consideration: the removal of CR terms from $\hat{H}_{chain}$ prevents the spontaneous creation of excitations in the system. In terms of the equivalent all-optical setups, this means that by preparing the register in a classical initial state, no inter-oscillator entanglement has to be expected, in this case. Indeed, in ref.~\cite{myungBS} it is shown that non-classicality at the inputs of a BS is a fundamental pre-requisite for the entanglement of its outputs. In presence of CR terms, the single-oscillator squeezing provides the necessary non-classicality for inter-oscillator entanglement. We will come back to this point later, when the entanglement generation is quantitatively addressed.

It is worth comparing the decomposition in Eq.~(\ref{decomposizione}) with what has been found for a star-shaped bosonic configuration~\cite{iohelenmyung}. In an open linear chain, $\hat{B}_{coll}$ induces multi-body interactions between the element of the chain. In particular, from Fig.~\ref{pattern} we see that an exchange of information is always required between the first and last element in a chain, a feature which holds regardless of $N$~\cite{commento2}. In the star-shaped configuration, on the other hand, any exchange of information occurs via a preferential way passing through the central component~\cite{iohelenmyung}.

\section{Entanglement in an open chain: symmetry of the variance matrix} 
\label{entanglement}

In order to investigate the dynamics of the entanglement generated among the oscillators in an open chain, we concentrate on Gaussian states and rely on the powerful tools provided by the variance matrix formalism. Indeed, the statistical properties of a Gaussian state, {\it i.e.} a state whose characteristic function is Gaussian, are entirely specified by the knowledge of its variance matrix. The variance matrix ${\bf V}$ is defined as $V_{\alpha\beta}=\langle\{\hat{x}_{\alpha},\hat{x}_{\beta}\}\rangle\,(\alpha,\beta=1,..,2N)$, where $\hat{x}=\hat{q},\hat{p}$ and, for convenience, we have adopted the ordering of the quadrature operators $\hat{\bf x}=(\hat{q}_{1},\hat{p}_{1},..,\hat{q}_{N},\hat{p}_{N})^{T}$. 
Throughout the paper, we assume that the Gaussian peak of each oscillator is at the origin of the respective phase-space. ${\bf V}$ is in one-to-one correspondence with the characteristic function of a Gaussian CV state which, in turns, gives information about the state of the system~\cite{myungmunro}. 

When applied to an $N$-oscillator input Gaussian state, the operations involved in Eq.~(\ref{decomposizione}) give an output state which is also Gaussian. They can be formally described by means of the transformations ${\bf {\cal T}}_{{\bf R}_{j}}(\phi_{j})=\cos\phi_{j}\one+i\sin\phi_{j}{\bm \sigma}_{y}$ for single-oscillator rotation and ${\bf {\cal T}}_{{\bf S}_{j}}(s_j)=e^{-s_{j}{\bm \sigma}_{z}}$ for single-oscillator squeezing, where ${\bm \sigma}_{\alpha}$ ($\alpha=y,z$) is the $\alpha$-Pauli matrix. For two-oscillator BS we have 
\begin{equation}
\label{transformation}
{\bf {\cal T}}_{{\bf B}_{jk}}(r_{jk},t_{jk})=
\begin{pmatrix}
t_{jk}\one&-r_{jk}\one\\
r_{jk}\one&t_{jk}\one
\end{pmatrix},
\end{equation}
where $t_{jk},\,r_{jk}$ stand for the transmittivity and reflectivity of the BS acting on elements $j$ and $k$ (with $t^{2}_{jk}+r^{2}_{jk}=1$)~\cite{myungmunro}. Explicitly, these transformations change an input variance matrix ${\bf V}$ to ${\bf V}_{\alpha_{j}}={\bf {\cal T}}^{T}_{\alpha_{j}}{\bf V}{\bf {\cal T}}_{\alpha_{j}}$ ($\alpha={\bf S},{\bf R}$) for a single-oscillator $2\times{2}$ variance matrix and ${\bf V}_{{\bf B}_{jk}}={\bf {\cal T}}^{T}_{{\bf B}_{jk}}{\bf V}{\bf {\cal T}}_{{\bf B}_{jk}}$ for a two-oscillator $4\times{4}$ variance matrix. From now on, we indicate with ${\bf V}^{N}_{f}$ the final variance matrix resulting from the application of all the transformations involved in $\hat{U}(t)$ for a given $N$. In this Section we focus the attention onto the case in which all the oscillators are prepared in vacuum state, so that the initial variance matrix of the joint state of the chain is ${\bf V}=\oplus^{N}_{j=1}\one_{j}$. By using Eq.~(\ref{decomposizione}) it can be shown that, for $N=3$, the final variance matrix reads
\begin{equation}
\label{VMB3}
{\bf V}^{3}_{f}=
\begin{pmatrix}
{\bf L}_{1}&{\bf C}_{12}&{\bf C}_{13}\\
{\bf C}^{T}_{12}&{\bf L}_{2}&{\bf C}_{12}\\
{\bf C}^{T}_{13}&{\bf C}^{T}_{12}&{\bf L}_{1}
\end{pmatrix},
\end{equation}
where ${\bf L}_{1}=\one+{\bf C}_{13}$ and ${\bf L}_{2}=\one+2{\bf C}_{13}$ account for the local properties of the oscillators while ${\bf C}_{12}=\frac{1}{\sqrt{2}}\sum^{3}_{j=2}(-1)^{j+1}{\bf c}_{j}$ and ${\bf C}_{13}=\frac{1}{2}\sum^{3}_{j=2}{\bf c}_{j}$ describe the inter-oscillator correlations. We have introduced the elementary correlation matrices (which depend on the effective squeezing and rotations of oscillators $j=2,3$)
\begin{equation}
\label{c}
{\bf c}_{j}=
\begin{pmatrix}
-e^{-2s_{j}}\sin^{2}(\phi_{j})\sinh(2s_{j})&\frac{1}{2}\sin(2\phi_{j})\sinh(2s_{j})\\
\frac{1}{2}\sin(2\phi_{j})\sinh(2s_{j})&e^{2s_{j}}\sin^{2}({\phi}_{j})\sinh(2s_{j})\\
\end{pmatrix}.
\end{equation}
It is remarkable in Eq.~(\ref{VMB3}) that the oscillators $1$ and $3$ have the same local properties, which are different from those of the mediator oscillator $2$. Moreover, the correlations between oscillators $1$ and $2$ appear to be the same as those between $2$ and $3$, which witnesses an evident degree of symmetry in the bosonic system ruled by Eq.~(\ref{interazione}). The proportionality of the correlation matrix ${\bf C}_{12}$ to the difference ${\bf c}_{3}-{\bf c}_{2}$ is important, in this analysis, and is in striking contrast with the inherent structure of the correlations between the end points of the chain. These observations will be crucial in the upcoming discussion relative to the improvement of the end-to-end entanglement. The structure of Eq.~(\ref{VMB3}) is found to hold for larger registers. Indeed, as still manageable examples, we mention that for $N=4$ and $5$ the decomposition of $\hat{U}(t)$ is such that  
\begin{equation}
\label{VMB5}
{\bf V}^{4}_{f}=
\begin{pmatrix}
{\bf L}_{1}&{\bf C}_{12}&{\bf C}_{13}&{\bf C}_{14}\\
{\bf C}^{T}_{12}&{\bf L}_{2}&{\bf C}_{23}&{\bf C}_{13}\\
{\bf C}^{T}_{13}&{\bf C}^{T}_{23}&{\bf L}_{2}&{\bf C}_{12}\\
{\bf C}^{T}_{14}&{\bf C}^{T}_{13}&{\bf C}^{T}_{12}&{\bf L}_{1}
\end{pmatrix},
\end{equation}
which extends the symmetry already manifested in ${\bf V}^{3}_{f}$. In fact, the symmetry is a general property of ${\bf V}^{N}_{f}$: it is straightforward to see that ${\bf V}^{5}_{f}$ exhibits symmetry with respect to the central element of the chain, whose local properties are unique in the system. The expressions of ${\bf C}_{jk}$'s in terms of elementary correlation matrices analogous to ${\bf c}_{j}$ in Eq.~(\ref{c}) are, in general, quite cumbersome. 

We address the generation of quantum correlations among the elements of an $N$-oscillator open chain 
as well as a simple strategy suitable for the improvement of the performances of this bosonic system as a long-haul entanglement distributor. The Gaussian preserving nature of the transformations ${\bf {\cal T}}_{{\bf R}_{j}}(\phi_{j})$, ${\bf {\cal T}}_{{\bf S}_{j}}(s_j)$ and ${\bf {\cal T}}_{{\bf B}_{jk}}(r_{jk},t_{jk})$ allows us to exploit the well-known necessary and sufficient conditions for the entanglement of two-body CV Gaussian states~\cite{simon,myungmunro}. The explicit object of our investigation will be the evaluation of the bipartite entanglement between the first and the $j$-th oscillator in a chain of $N$ oscillators. Therefore, we consider the reduced variance matrices ${\bf\it v}_{1j}$ of the pairs $(1,j)$ which are found from ${\bf V}_{f}$ by extracting the $4\times{4}$ submatrices ($j=2,..,N$)
\begin{equation}
{\bf\it v}_{1j}=
\begin{pmatrix}
V_{1,1}&V_{1,2}&V_{1,2j-1}&V_{1,2j}\\V_{2,1}&V_{2,2}&V_{2,2j-1}&V_{2,2j}\\V_{2j-1,1}&V_{2j-1,2}&V_{2j-1,2j-1}&V_{2j-1,2j}\\V_{2j,1}&V_{2j,2}&V_{2j,2j-1}&V_{2j,2j}
\end{pmatrix}.
\end{equation} 
As a measure of entanglement we use the {\it logarithmic negativity} which provides an upper bound to the entanglement of distillation~\cite{vidalwerner} and is strictly related to the extent to which a given state violates the Peres-Horodecki criterion for separability~\cite{npt}. For bipartite Gaussian states, this entanglement measure can be easily calculated starting from the symplectic spectrum of the partial transposition of the variance matrix ${\bf{\it v}}_{ab}$. In the phase-space, the partial transposition with respect to oscillator $b$ corresponds to the time-reversal operation which flips the sign of the momenutm quadrature operator of $b$. This can be represented by the action of the matrix $P=\one\oplus{\bm \sigma}_{z}$ onto ${\bf\it v}_{ab}$. We introduce the matrix $\Sigma_{ab}=\Sigma_{a}\oplus\Sigma_{b}$, where $\Sigma_{m}=i{\bm \sigma}_{y,m}$ ($m=a,b$) is the symplectic matrix of oscillator $m$~\cite{simon}. The symplectic eigen-values of ${\bf{\it v}}'_{ab}=P{\bf{\it v}}_{ab}P$ are the eigen-values of $\modul{i\Sigma_{ab}{\bf{\it v}}'_{ab}}$, which are always equal in pairs. By calling $\gamma'_{n}$ ($n=1,2$) the representative of each pair, the inequality $\min_{n}({\gamma'_{n}})\ge{1}$ is a necessary and sufficient condition for the separability of ${\bf\it v}_{ab}$. The logarithmic negativity $\Lambda^{ab}_{ng}$ is then evaluated as $\Lambda^{ab}_{ng}=\sum_{n}\max\left(0,-\log_{2}{\gamma'_{n}}\right)$~\cite{vidalwerner}. 

\begin{figure}[t]
\centerline{\psfig{figure=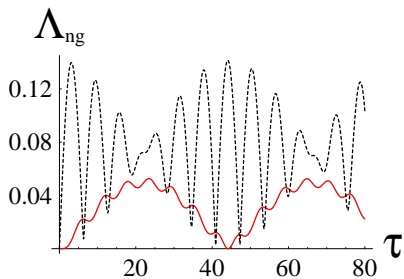,width=6.0cm,height=3.7cm}}
\caption{(color online) Logarithmic negativity $\Lambda^{1j}_{ng}$ ($j=2,3$), for a three-element open chain, plotted against the dimensionless interaction time $\tau=\omega{t}$, for $\kappa/\omega=0.1$. The dotted line represents the behavior of $\Lambda^{12}_{ng}=\Lambda^{23}_{ng}$, the solid line $\Lambda^{13}_{ng}$.} 
\label{ent3NSQ}
\end{figure}

The calculation of $\Lambda^{1j}_{ng}$ for $N=3$, $j\in[2,3]$ and $\kappa/\omega=0.1$ leads to the plots shown in Fig.~\ref{ent3NSQ}, where the bipartite entanglement between the three oscillators in the open chain is plotted against the rescaled interaction time $\tau=\omega{t}$. The choice for the ratio $\kappa/\omega$ is dictated by the fact that, experimentally, a weak coupling regime of $\kappa\ll\omega$ is the only realistic situation \cite{plenio,iohelenmyung}. 
\begin{figure}[b]
\hskip1cm{\bf (a)}\hskip4cm{\bf (b)}
\psfig{figure=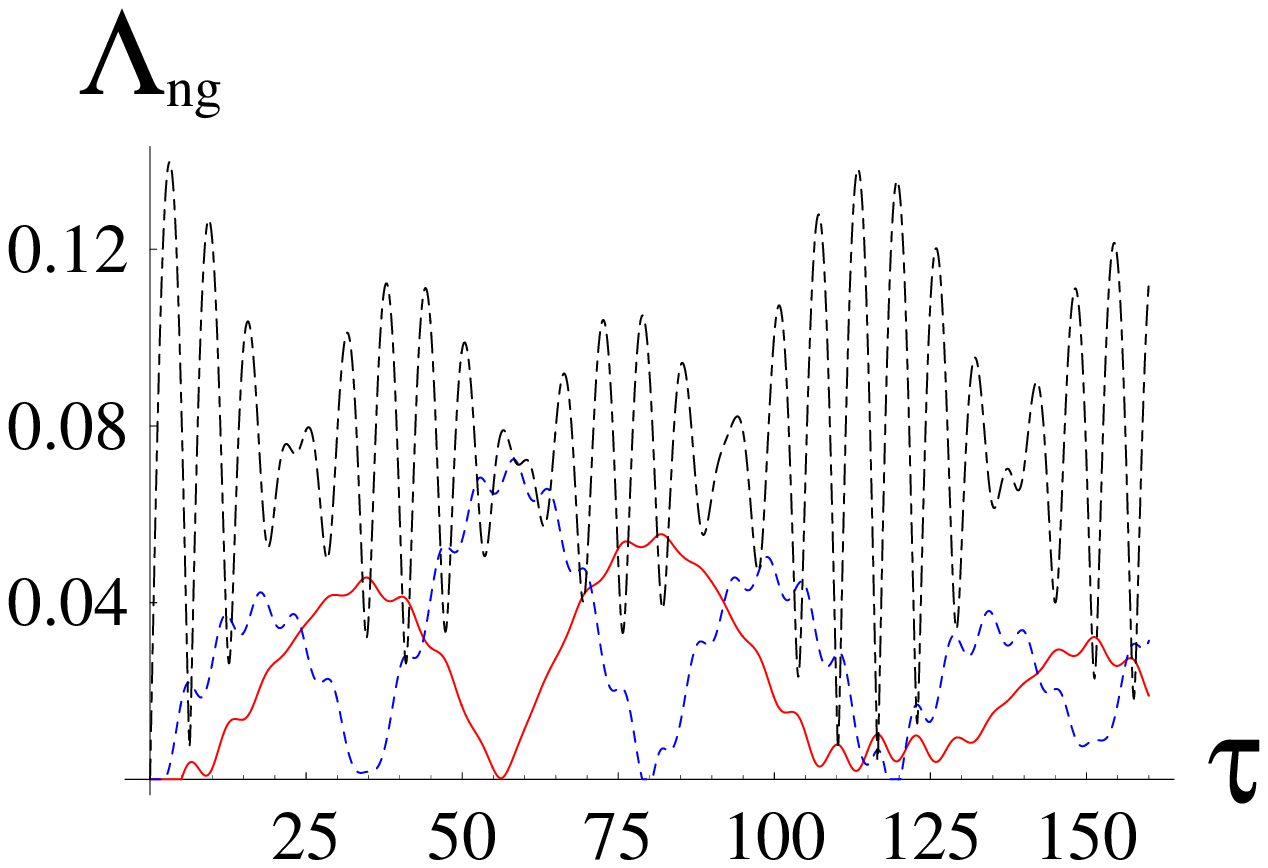,width=4.6cm,height=3.2cm}\psfig{figure=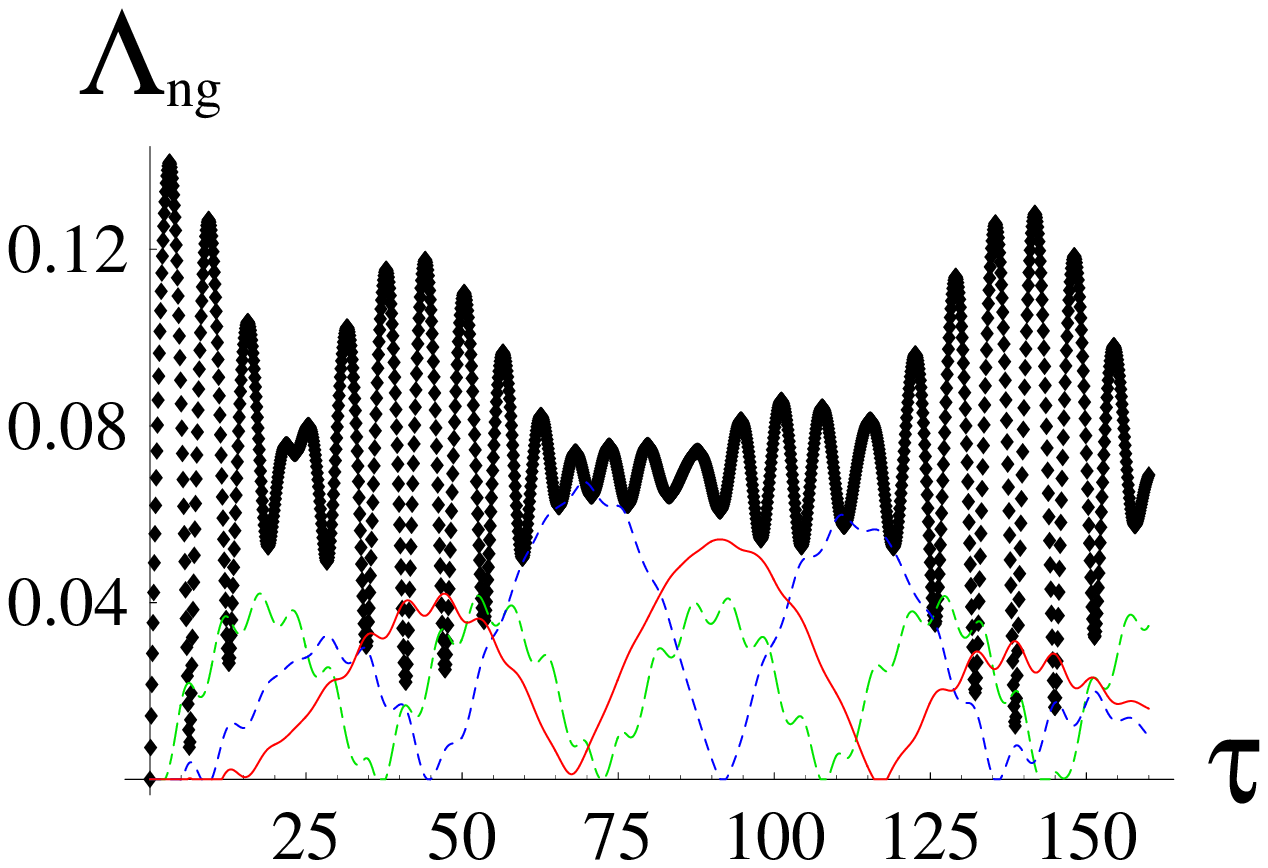,width=4.6cm,height=3.2cm}
\caption{(color online) {\bf (a)}: Logarithmic negativity $\Lambda^{1j}_{ng}$ ($j=2,3,4$), for a four-element open chain, plotted against the dimensionless interaction time $\tau=\omega{t}$, for $\kappa/\omega=0.1$. The dot-dashed line is for $\Lambda^{12}_{ng}$, the dotted line is for $\Lambda^{13}_{ng}$ and, finally, the solid line represents $\Lambda^{14}_{ng}$. {\bf (b)}: Same as panel {\bf (a)} but for $N=5$. The thick dotted line is for $\Lambda^{12}_{ng}$, the thin dot-dashed curve is for $\Lambda^{14}_{ng}$ and the dashed one is for $\Lambda^{13}_{ng}$. Finally, the solid line shows $\Lambda^{15}_{ng}$.} 
\label{ent4NSQ}
\end{figure}
The periodic behavior of the functions plotted is the signature of the time-dependence of the collective rotation $\hat{R}_{coll}(t)$. As seen in the symmetry of ${\bf V}^{3}_{f}$, we have $\Lambda^{12}_{ng}=\Lambda^{23}_{ng}$ (Fig. \ref{ent3NSQ}, dotted curve). The peak of $\Lambda^{12}_{ng}$ occurring at $\tau=\tau^{*}\simeq{44.2}$ corresponds to $\modul{\sin\!{\phi_{j}}}>{0.998}$. At this instant of time all the off-diagonal elements of ${\bf c}_{j}$'s do not exceed $\sim5\times{10}^{-3}$. On the other hand, it is evident that $\Lambda^{13}_{ng}$ (Fig.~\ref{ent3NSQ}, solid line) is smaller than $\Lambda^{12}_{ng}$ practically for any value of $\tau$, entirely disappearing at $\tau^{*}$. Thus, despite the CR terms are responsible for the {\it for free} generation of entanglement, a {\it passive} approach in which the bosonic register evolves freely without external intervention is evidently unsuitable for the creation of a reliable end-to-end entangled channel. On the contrary, almost all the quantum correlations within the system are localized among the nearest-neighbor oscillators (subsystems $1+2$ and $2+3$). The trend is common to any other case we have checked: $\Lambda^{12}_{ng}$ can be almost an order of magnitude larger than any other $\Lambda^{1j}_{ng}$ (see Fig.~\ref{ent4NSQ}, for example). Moreover, it is apparent that the behavior of each entanglement function persists by enlarging the register. Only small modifications are observed in $\Lambda^{1j}_{ng}$ when $N\rightarrow{N}+1$, the most evident of which is that $\Lambda^{1N}_{ng}$ becomes non-zero after an increasing time delay. This can be understood by considering the effective all-optical setup (as the one sketched in Fig.~\ref{3oscillatori}): when all the oscillators are initially prepared in their vacuum state, the first set of BS's (on the left hand side of the figure) are ineffective as they superimpose $\ket{0}$ states. As soon as the squeezing of oscillators $2,..,N$ is performed, the second set of BS's, on the right hand side of the figures, is responsible for the generation and {\it propagation} of quantum correlations. Obviously, the number of operations which precede the coupling between $1$ and $N$, this latter carrying all the necessary non-classicality, increases with the dimension of the register, thus retarding the settlement of their entanglement. Again, the decomposition Eq.~(\ref{decomposizione}) shades new light onto the important features of the entanglement dynamics throughout the system, complementing the results highlighted by previous analyses~\cite{plenio}. 

\begin{figure}[b]
\centerline{\psfig{figure=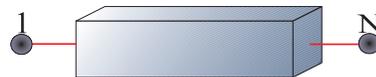,width=5.0cm,height=1.0cm}}
\caption{(color online) Scheme of principle for the end-to-end entanglement generation procedure. The two-terminal device embodies the bosonic quantum channel over which we assume to have no local control. On the other end, $1$ and $N$ are assumed to be within the grasp of two parties, which can arbitrarily prepare and measure the state of the respective oscillator. The interaction which determines the set of $\kappa$'s globally addresses the $N$-element system.} 
\label{principio}
\end{figure}

\section{Tagging by squeezing}
\label{improving}

For the purpose of creating an entangled state of the extremal oscillators in the channel, it is certainly desirable to look for strategies which quantitatively improve the entanglement settled between $1$ and $N$. Moreover, we would like to find out a way to make $\Lambda^{1N}_{ng}$ dominant with respect to any other $\Lambda^{1j}_{ng}$. The approach we are going to follow does not rely on local control over the elements of the chain between the first and the last oscillators. We assume that $1$ and $N$ are held by two spatially separated parties who can perfectly control the preparation of the respective oscillator and, if required, can measure their state. On the other hand, the interactions between the oscillators in the chain are set in a global way by a potential which collectively addresses all the elements at the same time. This approach is entirely within the rules of the {\it global addressing} strategies exemplified by quantum state transfer and phase-covariant cloning in quantum spin chains~\cite{qst,cloning} and by {\it always-on} computational schemes~\cite{sougato}. In this perspective, the chain is seen as a two-terminal device whose intermediate stage is embodied by the $N-2$ oscillators between the {\it ending} terminals $1$ and $N$. This central section is a black box whose dynamics are out of the grasp.

Intuitively, one would like to magnify the inherent distinction of the pair of oscillators $1$ and $N$, shown by Eqs.~(\ref{VMB3}) and~(\ref{VMB5}), from the rest of the register. Therefore, any local action performed onto the ending terminals of the device in Fig.~\ref{principio}, has to be designed so that the local properties of oscillators $1$ and $N$ still remain mutually equal. By considering the analysis performed in Section~\ref{modello} and the role that non-classical states have in the entanglement by means of coupler operators~\cite{myungBS}, we look for an initial preparation of the register which can result in a quantitative increase of the end-to-end degree of entanglement. 
After a close inspection of the decomposition, we conjecture that single-oscillator squeezing operations onto $1$ and $N$ 
should improve the degree of entanglement between them. 

\begin{figure}[b]
{\bf (a)}\hskip3cm{\bf (b)}
\psfig{figure=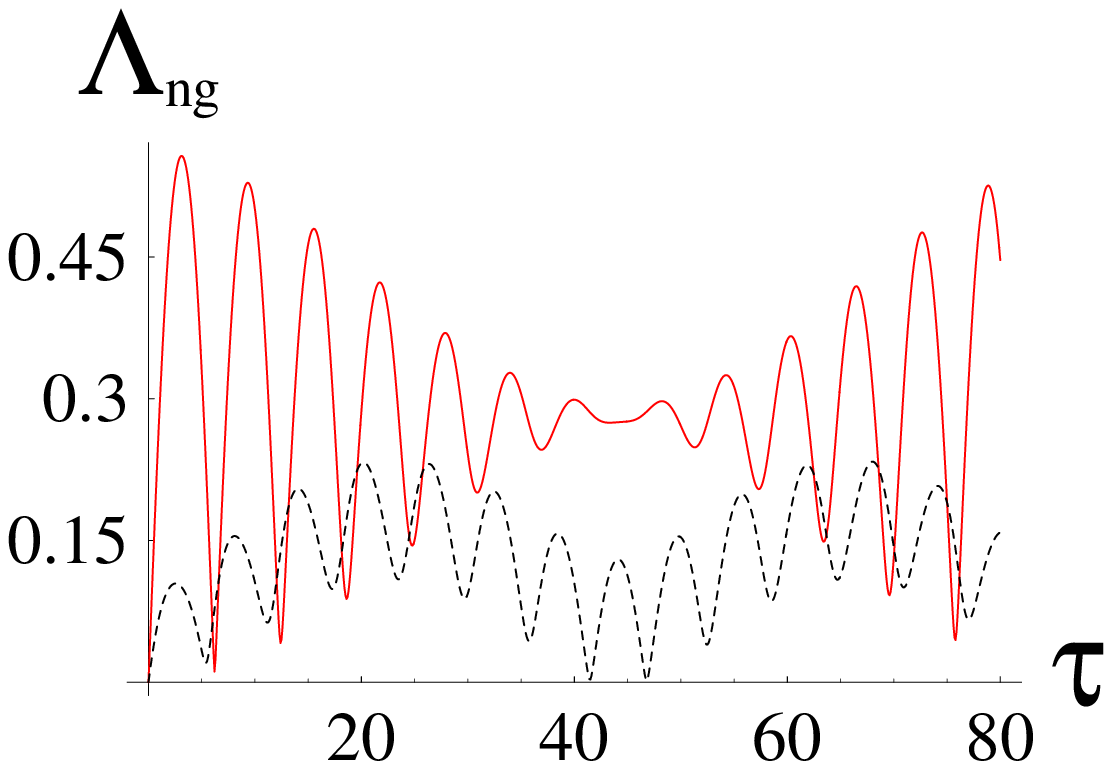,width=4.6cm,height=3.2cm}\psfig{figure=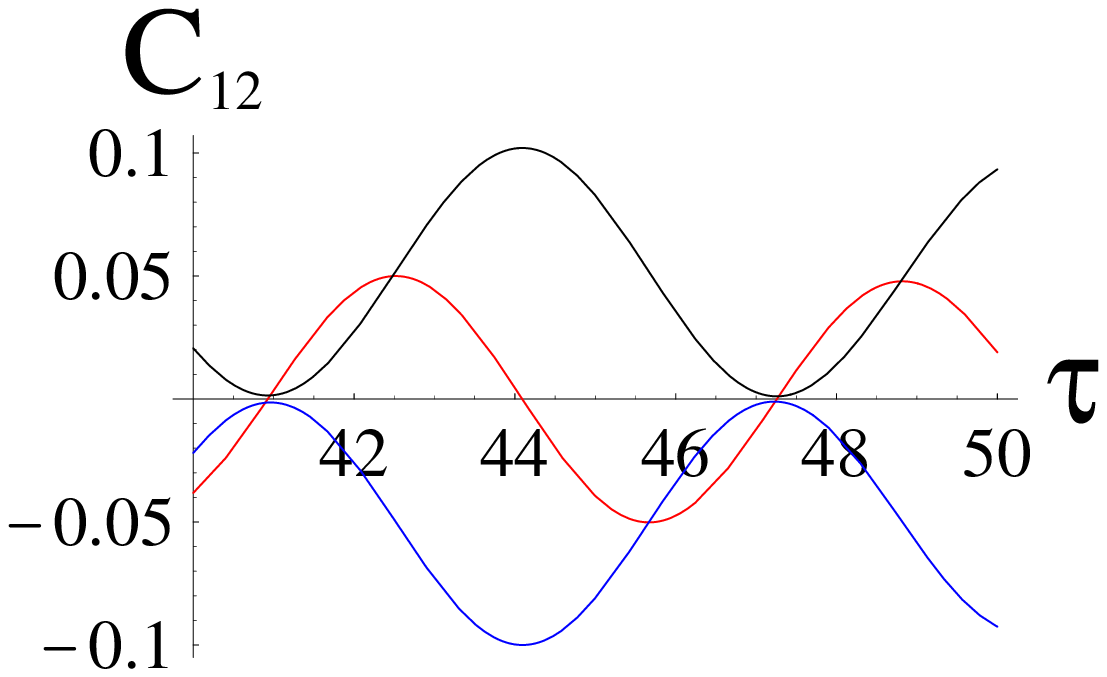,width=4.6cm,height=3.2cm}
\caption{(color online) {\bf (a)}: Logarithmic negativity $\Lambda^{1j,tag}_{ng}$ ($j=2,3$) for a {\it tagged} three-oscillator open chain, plotted against the dimensionless interaction time $\tau=\omega{t}$, for $\kappa/\omega=0.1$ and $r=0.2$. The dashed line is for $\Lambda^{12,tag}_{ng}=\Lambda^{23,tag}_{ng}$ while the solid line shows $\Lambda^{13,tag}_{ng}$. {\bf (b)}: Time behavior of the elements of the correlation matrix ${\bf C}_{12}$ for the tagged chain of panel ${\bf (a)}$.} 
\label{ent3SQ}
\end{figure}

In order to demonstrate our conejcture, we address the case of $N=3$ and we consider the preparation of an initial state whose variance matrix reads ${\bf V}={\bf V}_{sq,1}\oplus\one_{2}\oplus{\bf V}_{sq,3}$, where ${\bf V}_{sq,j}=e^{-2r{\bm \sigma}_{z,j}}$ is the variance matrix of a squeezed state with its squeezing parameter $r$. The calculation of the logarithmic negativity for the subsystems $1+2,\,1+3$ and $2+3$ can proceed according to the recipe given in Section~\ref{entanglement}. The corresponding degree of entanglement, from now on, will be indicated as $\Lambda^{jk,tag}_{ng}$ ($j,k\in[1,N]$). The results, for $r=0.2$, are shown in Fig.~\ref{ent3SQ} {\bf (a)}.
Due to the symmetry of the particular initial preparation, it is easy to check that we get a final variance matrix having the same general structure as Eq.~(\ref{VMB3}), with suitably modified elementary correlation matrices. In particular, Fig.~\ref{ent3SQ} ${\bf (b)}$ shows the time behavior of the matrix elements of ${\bf C}_{12}={\bf C}_{23}$. Differently from what happens for an initially prepared vacuum state, when the ending elements are initially squeezed and for $\tau\simeq46.8$, all the elements $(C_{12})_{jk}$ simultaneously become very close to zero ($\modul{(C_{12})_{1,1}}\simeq6\times10^{-3},\,\modul{(C_{12})_{2,2}}\simeq4\times10^{-3}$ with $\modul{(C_{12})_{1,2}}=\modul{(C_{12})_{2,1}}=0$), whereas ${\bf C}_{13}$ (at that value of $\tau$) becomes diagonal with matrix elements in the range of $0.1$. This accounts for the improvement of the entanglement settled between $1$ and $3$ with, correspondingly, $\Lambda^{12,tag}_{ng}<2\times10^{-3}$. The subsystem $1+3$ has been {\it tagged} by the single-element pre-squeezing to be the preferential pair of oscillators for the entanglement generation within the chain. It is worth stressing that this {\it tagging} procedure is possible in virtue of the symmetry existing between the ending elements of the open chain. An analogous conclusion has been drawn in ref.~\cite{salerno}, where a {\it totally symmetric} $N$-body CV system has been considered in order to point out the possibility of a unitary localization of the entanglement. In our case, however, the CV chain exhibits a degree of symmetry which is inferior to the one treated in ref.~\cite{salerno}. Different pairs of oscillators are characterized by different local and correlation properties, which makes the problem approached in this paper intrinsically different from the one in~\cite{salerno}. Nevertheless, we have shown that entanglement localization is possible with a lower degree of symmetry, which is {\it per se} an interesting point.

\begin{figure}[b]
\hskip1cm{\bf (a)}\hskip4cm{\bf (b)}
\psfig{figure=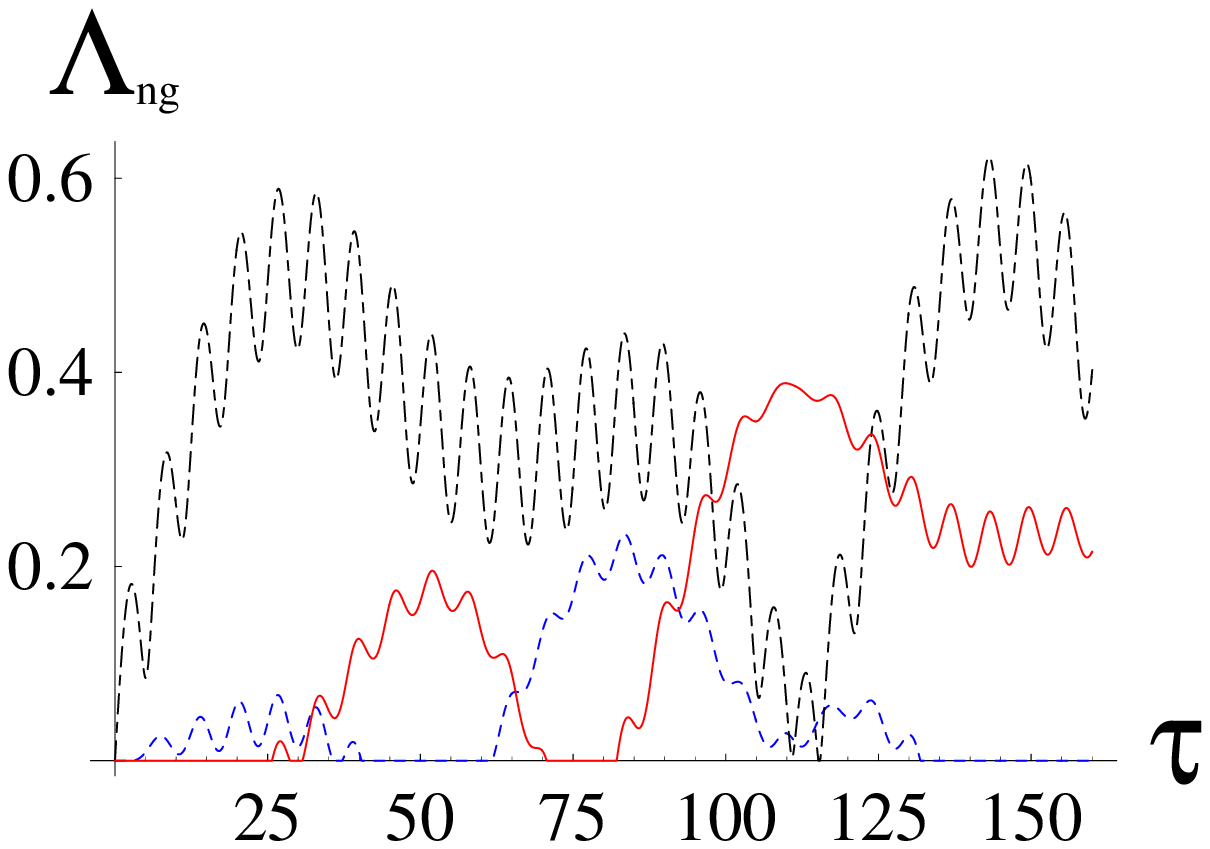,width=4.6cm,height=3.2cm}\psfig{figure=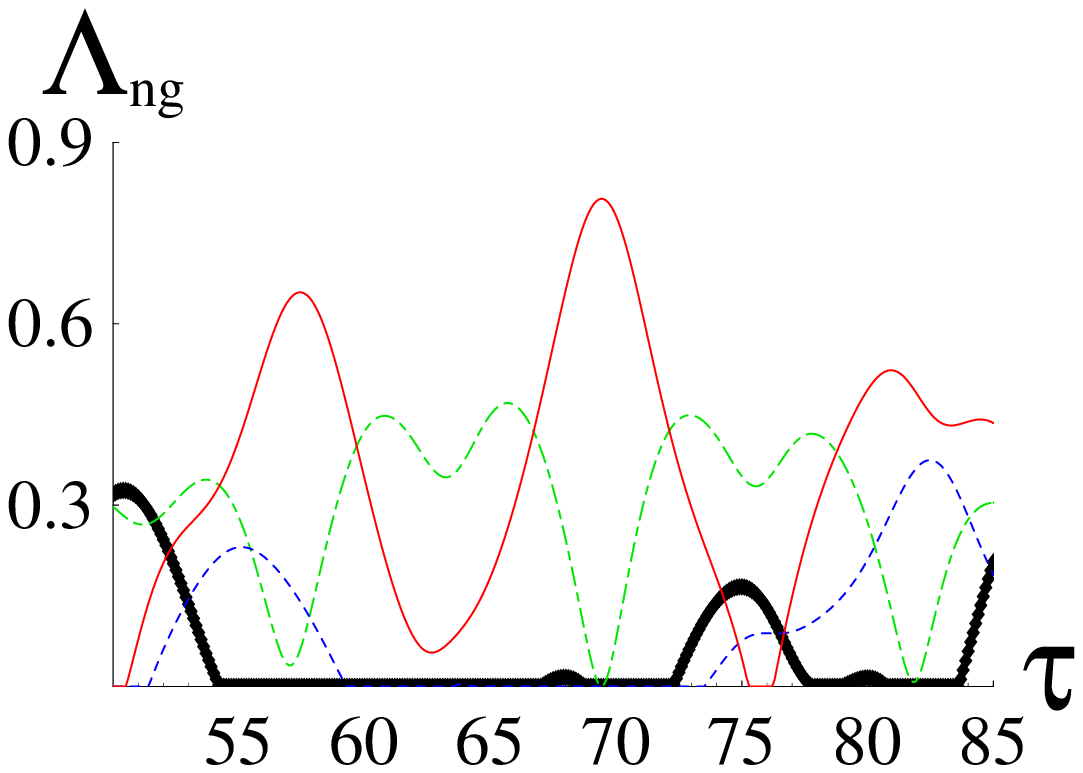,width=4.6cm,height=3.2cm}
\caption{(color online) {\bf (a)}: Logarithmic negativity $\Lambda^{1j,tag}_{ng}$ ($j=2,3,4$) for a tagged four-oscillator chain against $\tau=\omega{t}$, for $\kappa/\omega=0.1$. The dot-dashed line is for $\Lambda^{12,tag}_{ng}$, the dotted line is for $\Lambda^{13,tag}_{ng}$ and, finally, the solid line represents $\Lambda^{14,tag}_{ng}$. In this panel, $r=0.4$. {\bf (b)}: Same as panel {\bf (a)} but for $N=5$. The thick black line is for $\Lambda^{12,tag}_{ng}$, the thin dot-dashed curve is for $\Lambda^{14,tag}_{ng}$ and the dashed one is for $\Lambda^{13,tag}_{ng}$. Finally, the solid line shows $\Lambda^{15,tag}_{ng}$. In this panel, $r=0.6$.} 
\label{ent4SQ}
\end{figure}

We can generalize the choice for the initial variance matrix in the tagging procedure to the case of $N$ oscillators by considering ${\bf V}={\bf V}_{sq,1}\oplus\left(\oplus^{N-1}_{j=2}\one_{j}\right)\oplus{\bf V}_{sq,N}$. Again, an explicit calculation for the logarithmic negativity can be performed, which leads to the plots shown in Figs.~\ref{ent4SQ} ${\bf (a)}$ and ${\bf (b)}$, for the cases $N=4$ and $5$. In Fig.~\ref{ent4SQ} {\bf (b)}, the time-range has been restricted to the interesting region where $\Lambda^{15,tag}_{ng}\gg\Lambda^{1j,tag}_{ng},\,j\in[2,4]$ in order to make the plot more transparent.
It is evident that there is always at least one value of $\tau$ at which the end-to-end entanglement dominates, making the tagging procedure effective. The amount of pre-required single-oscillator squeezing slightly depends on the dimension of the register and the plots in this paper show those values of $r$ at which we have found a good trade-off between the degree of entanglement and the effectiveness of the tagging strategy. As we have stressed before, the even and odd cases are inherently different, as also witnessed by the fact that the value of $\tau$ corresponding to an optimized tagging operation is larger for $N=4$ than for $N=5$. At the same time, $\Lambda^{14,tag}_{ng}$, for $N=4$, is roughly proportional to $r$, while $\Lambda^{1N,tag}_{ng}>r$ for all the odd $N$ cases we have checked. 

\section{Remarks}

We have addressed the problem of long-haul entanglement creation in a register of bosons interacting via a global potential. The dynamics of entanglement can be clearly tracked via the effective decomposition of the time propagator in terms of simple linear optics elements as rotators, single-oscillator squeezers and couplers. This approach has allowed us to spot out a series of interesting features, characterizing the evolution of the quantum correlations settled among the elements of the register. As a result, we have been able to relate the conceptual role played by the CR terms in the entanglement generation process to effective squeezing operations on the elements of the register. The usefulness of this analysis is also witnessed by the design of a {tagging} protocol for the improvement of the end-to-end entanglement and the simultaneous reduction of any other $1\rightarrow{j}$ ($j=2,..,N-1$) quantum correlation in an chain of $N$ elements. No local control over the central section is required: a proper preparation of the extremal oscillators and a collective interaction are sufficient to achieve the task. We believe this formal approach could be used in order to clarify other aspects related to the role played by the CR terms in entanglement creation, an issue which is certainly relevant especially in many problems of solid-state physics.

\acknowledgments

We acknowledge discussions with Dr. J. Fiur\'a\v{s}ek. This work has been supported by the UK EPSRC and the Korea Research Foundation (2003-070-C00024).


\begin{thebibliography}{99}
\bibitem{commento} Differently from the odd number case, for the even one it is not easy to identify a suitable pattern of couplers for $N>4$. Despite this is not the focus of our study, it will be certainly interesting to investigate about the intrinsic effect of the parity onto the collective dynamics of the oscillators in a chain.
\bibitem{commento2} It is interesting to notice that the BS between oscillators $1$ and $N$ always occurs, in our decomposition, before any other inter-element mixing, despite these two oscillators are far apart and not directly connected. In a sense, the interaction between $1$ and $N$ is mediated by the intermediate elements as it happens between two electronic states of a multilevel atoms belonging to a dipole-forbidden transition. In this latter case, an effective exchange of population can be attained through interference induced by multiphoton couplings (for instance a two-photon Raman transition in a three-level atom).
\end{thebibliography}
\end{document}